\documentclass[runningheads]{llncs}

\usepackage[T1]{fontenc}
\usepackage{multirow}
\usepackage{subfig}
\usepackage{graphicx}
\usepackage[utf8]{inputenc} 
\usepackage[T1]{fontenc}    
\usepackage{hyperref}       
\usepackage{url}            
\usepackage{booktabs}       
\usepackage{amsfonts}       
\usepackage{nicefrac}       
\usepackage{microtype}      
\usepackage{xcolor}         
\usepackage{amsmath}
\usepackage{algorithm}
\usepackage{algpseudocode}
\usepackage{graphicx}
\usepackage{tikz}
\usepackage{graphicx}
\usepackage{array}  
\usetikzlibrary{shapes, arrows.meta, positioning}
\usepackage{ragged2e}
\justifying

\title{MEDDAP: Medical Dataset Enhancement via Diversified Augmentation Pipeline}
\begin{document}
%

\author{Yasamin Medghalchi{\inst{1}} \and 
Niloufar Zakariaei\inst{1,2} \and 
Arman Rahmim\inst{1,2,3,4} \and 
Ilker Hacihaliloglu\inst{3,5}
}

\authorrunning{Medghalchi et al.}

\institute{
School of Biomedical Engineering, University of British Columbia \and
Department of Integrative Oncology, BC Cancer Research Institute \and
Department of Radiology, University of British Columbia \and
Department of Physics, University of British Columbia\and
Department of Medicine, University of British Columbia \\
\email{yasimed@student.ubc.ca}, \email{nilouzk@student.ubc.ca}, \email{Arman.rahmim@ubc.ca}, \email{ilker.hacihaliloglu@ubc.ca}
}

\maketitle              

\begin{abstract}
The effectiveness of Deep Neural Networks (DNNs) heavily relies on the abundance and accuracy of available training data. However, collecting and annotating data on a large scale is often both costly and time-intensive, particularly in medical cases where practitioners are already occupied with their duties. Moreover, ensuring that the model remains robust across various scenarios of image capture is crucial in medical domains, especially when dealing with ultrasound images that vary based on the settings of different devices and the manual operation of the transducer.
To address this challenge, we introduce a novel pipeline called MEDDAP, which leverages Stable Diffusion (SD) models to augment existing small datasets by automatically generating new informative labeled samples. Pretrained checkpoints for SD are typically based on natural images, and training them for medical images requires significant GPU resources due to their heavy parameters. To overcome this challenge, we introduce USLoRA (Ultrasound Low-Rank Adaptation), a novel fine-tuning method tailored specifically for ultrasound applications. USLoRA allows for selective fine-tuning of weights within SD, requiring fewer than 0.1\% of parameters compared to fully fine-tuning only the UNet portion of SD.
To enhance dataset diversity, we incorporate different adjectives into the generation process prompts, thereby desensitizing the classifiers to intensity changes across different images. This approach is inspired by clinicians' decision-making processes regarding breast tumors, where tumor shape often plays a more crucial role than intensity. In conclusion, our pipeline not only outperforms classifiers trained on the original dataset but also demonstrates superior performance when encountering unseen datasets.

\end{abstract}
\section{Introduction}

Breast cancer diagnosis heavily relies on mammography as the primary imaging modality due to its widespread availability and established effectiveness in early detection. However, mammography has inherent limitations, including discomfort for some patients and reduced sensitivity in dense breast tissue\cite{lee2016digital}. As an alternative, ultrasound has emerged as a valuable supplementary tool in breast imaging. Ultrasound offers several advantages, such as improved comfort during imaging and free from radiation exposure. Moreover, it provides real-time imaging capabilities, making it suitable for further evaluation of suspicious findings detected on mammograms. However, the efficacy of ultrasound is often questioned due to its variable sensitivity and specificity, largely influenced by the operator's skill and the interpretative complexity of echogenic patterns \cite{svensson2010advanced}. In this context, the rise of computational methodologies, particularly deep learning, has been identified as a promising approach to mitigate these challenges. Deep learning's ability to process and learn from extensive datasets could significantly enhance the diagnostic accuracy of ultrasound imaging in breast cancer \cite{litjens2017survey}.

The potential of integrating deep learning into ultrasound imaging is especially significant for improving breast cancer care in resource-limited settings. Ultrasound's portability and cost-effectiveness, augmented by the advancements in point-of-care ultrasound technology, present a viable alternative in regions where mammography access is constrained \cite{lee2016digital}. Nevertheless, the application of deep learning in this domain is not without challenges. The accuracy of these models heavily relies on the quality and quantity of the data, where manual collection and the adjustment of image acquisition parameters can introduce significant variability, complicating the development of robust algorithms \cite{yap2017automated}. Moreover, the scarcity of high-quality, annotated breast ultrasound datasets poses a critical bottleneck, limiting the training and validation of deep learning models \cite{yap2017automated}. To address these challenges, our research introduces an innovative deep learning (DL) pipeline that leverages stable diffusion (SD) techniques for the augmentation and diversification of ultrasound datasets through prompt engineering \cite{rombach2021highresolution}. Previously, it was demonstrated that simply expanding the dataset volume is inadequate; a deliberate effort to diversify the dataset is essential to enhance diagnostic precision \cite{fujioka2020utility}. 

Accordingly, our research underscores the vital need for methodically enhancing and diversifying ultrasound datasets, to fully harness deep learning's diagnostic potential in breast cancer diagnosis. Our specific contributions include: \textbf{1-} A novel fine-tuning approach, termed Ultrasound Low-Rank Adaptation (USLoRA), integrating low-rank adaptation techniques into stable diffusion models, marking the first instance of such methodology within the realm of synthesizing ultrasound data. \textbf{2-} A text-based diffusion model guidance aimed at diversifying the generated ultrasound data, particularly focusing on enhancing the diversity of intensity distributions. \textbf{3-} Conducting extensive evaluation studies to demonstrate the diversity advantages of our proposed approach in improving classification accuracy.

\section{Related Works}
Variational Autoencoders (VAEs) and Generative Adversarial Networks (GANs) have been foundational in this space, with VAEs being lauded for their stability and ease of training by optimizing a lower bound on the log-likelihood, while GANs are recognized for their ability to produce high-quality images through adversarial training \cite{liang2022sketch,alsinan2020gan}. A combined style augmentation approach that incorporates information from various sources to enhance the training set was proposed in \cite{huang2023style}. However, conventional style transfers and traditional augmentation techniques fall short when addressing alterations in shape and they depend on the style image used \cite{zhao2022ood}. Recently, diffusion models have emerged as a potent force for producing synthetic three-dimensional images, presenting notable advancements over conventional generative models such as GANs \cite{dhariwal2021diffusion}. Their superior image synthesis capabilities, surpassing those of GANs, stem from their capacity to intricately refine details through a controlled and iterative refinement process. \cite{dhariwal2021diffusion,ho2020denoising,kazerouni2022diffusion,ali2022spot,khader2022medical}. Hence, the utilization of diffusion models in processing ultrasound data has garnered significant interest within the medical imaging community \cite{stojanovski2023echo,stevens2024dehazing,katakis2023generation}. These studies collectively illustrate the diverse potential of diffusion models in enhancing ultrasound diagnostics.  However, diffusion models often demand considerable computational resources, rendering them less viable in resource-constrained environments—an aspect largely overlooked by prior research in this domain. Furthermore, the exploration of generating diverse diffusion images with varying intensity distribution levels remains unaddressed in the context of ultrasound image analysis. 



\section{Method}

\subsection{USLoRA (Ultrasound Low-Rank Adaptation)}
We first explain the use of SD for generating synthetic images, denoted as $x^\prime_i$, using specific prompts $P$ that incorporate the class name $y^\prime_i$. In the SD process, an initial image $x$ undergoes encoding through $\mathcal{E}$, yielding $z = \mathcal{E}\{x\}$. This is followed by a forward process introducing Gaussian noise $\epsilon$ at each step $t$, resulting in $Z_t$. Subsequently, $Z_T$ and the Text condition $P$ are combined and input into the denoising U-Net, $\epsilon_\theta$, in the reverse process. The SD loss function is the mean squared error (mse), quantifying the difference between the actual noise $\epsilon$ and the noise predicted by $\epsilon_\theta(Z_t , t)$.

\begin{equation}
loss = {\mathbb{E}_{\mathcal{E}(x) , \epsilon \sim \mathcal{N}(0,\,I)}[||\epsilon - \epsilon_\theta(Z_t , t)||_2^2]}
\end{equation}

The utilization of SD is significantly hindered by its demanding GPU requirements, thus constraining its applicability within the bounds of existing computational resources. To address this challenge, we introduce the USLoRA (Ultrasound Low-Rank Adaptation) method. USLoRA represents the first application of LoRA (Low-Rank Adaptation)\cite{LORA} for fine-tuning SD specifically for ultrasound images. Conventional fine-tuning approaches necessitate updating the weight matrix, $W$, to align with the pre-trained model's weight, $W_0$. This process, inherently mimics the scale of the pre-trained weight. In standard fine-tuning, the weight matrix adjustment is represented as:

\begin{equation}
W = W_0 + \Delta W, \quad W \in \mathbb{R}^{d \times k}, \quad |W| = |W_0|
\end{equation}

However, Aghajanyan et al. have demonstrated that the weight change matrices are sparse and possess a low intrinsic rank\cite{intrinsic}. This insight gave rise to the LoRA method, which restructures the fine-tuning equation for large language models, like  GPT-3, to exploit the low intrinsic rank property\cite{LORA}.
In USLoRA the fine-tuning is performed through the product of two smaller matrices, $A$ and $B$, with rank $r$:
\begin{equation}
 W = W_0 + A B, \quad A \in \mathbb{R}^{d \times r}, \quad B \in \mathbb{R}^{r \times k} \quad r<< d
\end{equation}

In USLoRA, for fine-tuning SD for Ultrasound image generation, we harness the unique capabilities of LoRA \cite{LORA} specifically within the cross-attention layers of UNet architecture. The decision to freeze the remaining model components is strategic, focusing computational efforts precisely where they are most impactful - enhancing image representations guided by input prompts. These large models are typically trained on extensive datasets and demonstrate some capability in comprehending both images and prompts. Our objective is just to establish a connection between the provided prompts and images within our ultrasound dataset, necessitating fine-tuning of the cross-attention layers using LoRA for weight matrices ${ W_q, W_k, W_v, W_o }$. Finally, we fine-tune SD with less than 0.1\% parameters compared to fine-tuning the whole UNet module of SD.

\subsection{Prompt Engineering}
We utilize SD as a text-to-image generator, requiring textual descriptions as input. To enrich our dataset effectively, we enhance input prompts with descriptive adjectives. Our adjective selection is driven by two observations. First, we consider the variability of ultrasound image intensity arising from device differences and transducer orientation. Second, we aim to challenge classifiers with diverse clinical scenarios while recognizing that tumor detection relies more on structural features than intensity. Given the limited dataset size, we ensure these additions remain general rather than overly specific to avoid overwhelming the model with excessive detail. Our augmentation strategy includes adjectives such as 'colorful,' 'solarized,' and 'stylized'. Our prompt structure follows the format ' \{adj\} ultrasound image of {benign/malignant/no} tumor in the breast'. By integrating class labels (benign/malignant) into the SD model, we mitigate errors stemming from ambiguity in the input text. Moreover, the inclusion of adjectives enhances the diversification of SD model generations, capturing a wide array of discriminative features that can enhance classification accuracy.

\section{Experiments and Results}
\raggedright\textbf{Dataset.} In this study, we utilize three open-source breast ultrasound datasets. Dataset A was exclusively employed for training the SD model and the classifiers utilized to evaluate the enhancements attained with our augmented data for the downstream task. Dataset B was only utilized as test data for the downstream task of classification. 
\vspace{-\topsep}
\begin{itemize}
    \item[]\justify \textbf{Dataset A.} Comprising images from female patients aged 25 to 75, aimed at detecting breast cancer \cite{Dataset}, this dataset, collected in 2018, includes 780 ultrasound images from 600 patients, with an average size of 500x500 pixels. Categorized into normal, benign, and malignant classes, it consists of 133, 437, and 210 images, respectively. We partition the dataset into training (545 images), validation sets (115 images), and testing (120 images) at the patient level \cite{Dataset}. The training set is utilized for training and fine-tuning SD. Given the limited dataset size, we employ the five-fold cross-validation technique to conduct our experiments.
    \item[]\justify \textbf{Dataset B.} Dataset B, an open-access dataset available on Kaggle\cite{kaggle}, includes ultrasound images depicting benign and malignant breast cancers. These images have undergone augmentation via rotation and sharpening to augment the dataset size. For testing the trained classifiers, we utilize the test set from this dataset, comprising 500 images for benign cases and 400 images for malignant cases \cite{Dataset_c}.
\end{itemize}

\justify\textbf{Implementation Details.} We based our experiments on a pre-trained model\cite{compvis2023stable}, initialized with Stable-Diffusion-v-1-2 checkpoint weights and fine-tuned for 225k steps at 512x512 resolution on "laion-aesthetics v2 5+". We conducted experiments to fine-tune the proposed USLoRA method using hyperparameters of 2, 4, and 8. Initially, we generated images equivalent in size to the training set and compared them based on the FID score \cite{FID}. The resulting FID scores were 0.463, 0.357, and 0.513 for ranks of 2, 4, and 8, respectively. Consequently, we determined that a rank of 4 yields the optimal outcome. Training USLoRA on SD for Dataset A is done with 100 epochs of AdamW optimizer with a learning rate of $10^{-4}$, and the batch size is 1, requiring 6.9 GB of GPU memory and less than 5 hours of one NVIDIA GeForce RTX 2080 Ti GPU. The input image dimensions were set to $224 \times 224$. To demonstrate the utility of synthesized ultrasound images in enhancing the performance of a supervised downstream classification task, we leverage ResNet34 \cite{resnet}, SqueezeNet1.1 \cite{SqueezeNet}, and DenseNet121 \cite{densenet} architectures, widely recognized as baseline models in medical imaging. The classifiers were trained for 100 epochs with a batch size of 32. In our experimental setup, we employed the Adam optimizer for its efficiency in handling sparse gradients and its adaptability to large-scale data problems. The learning rate was set to 0.001, following conventional best practices for achieving a balance between convergence speed and training stability. It's important to note that while these architectures are integral to our evaluation, they do not represent the central focus of our study. Instead, our primary contribution lies in the synthesis of images, which can augment a range of CNN-based network architectures. \\
\textbf{Effect of adjectives in classification performance.}
In this section, we commence by training the proposed text-guided SD model on the train set of Dataset A. Notably, the generation process is confined to the train set of Dataset A exclusively, with no utilization of images from the validation or test set of Dataset A during SD model training. 

\begin{table*}[h]
    \caption{Analyzing the impact of various adjectives with different extension ratios on classifer accuracy. "" means SD was trained without an adjective. \% indicates the size of the synthesized data included in the training data. Test data comprised real ultrasound data from Dataset A. A lower FID indicates better-quality images. \textbf{Bold text indicates the best results achieved}.}
    \centering
    \begin{tabular}{lc|ccc|ccc|ccc}
        & {FID}
        & \multicolumn{3}{c}{Densenet121} & \multicolumn{3}{c}{ResNet34} & \multicolumn{3}{c}{Squeezenet1.1} \\
        Adjective & &50\% & 100\% & 200\% & 50\% & 100\%& 200\% & 50\% & 100\%& 200\% \\
        \midrule
        "" &0.357 &86.66 &80.83 & 80.83
        &85 & 81.66 & 83.33
        &78.33 & 69.17 &75 \\
        "Colorful" & 0.374&86.66 & 86.66 & 83.33
        &85 &81.66 &79.17
        &80 &68.33 &71.67\\
        "Stylized" &0.367 &83.33 & 87.5 & 81.67
        & \textbf{87.5} & 80 &81.67
        &76.67 & 78.33 &75 \\
        "High-contrast" & 0.561&\textbf{90} & 83.33 &83.33 
        &85 & 76.66 & 83.33
        &\textbf{82.5} &81.67&70.83\\
        "Low-contrast" & 0.503&84.16 &82.5 & 80.83
        &85.83 &79.16 &77.5
        &67.5 & 72.5 &77.5\\
        "Posterized" &0.435 &80.83 & 87.5 &86.67
        &85.83 & 80 & 79.17
        &69.17 &78.33 &74.17\\
        "Sheared" &0.499 &83.33 &87.5 & 84.17
        &83.33 &85 &76.67
        &75.83 & 67.5 &70.83\\
        "Solarized" &\textbf{0.26} &86.66 &85 &83.33
        &\textbf{87.5} &83.33 &84.17
        &74.17 & 77.5 &73.33\\
        "Bright" &0.663 &84.16 &83.33 &84.17
        &84.16 & 75 &77.5
        &74.17 &73.33 &73.33\\
        "Dark" & 0.354&85.83 &87.5 & 82.5
        &76.66&78.33 &78.33
        &77.5 & 74.17 &72.5\\
    \end{tabular}
    \label{table_adjectives}
\end{table*}

\begin{table*}[h]
\caption{Comparison of classification performance between classifiers trained solely on real data and those trained on a combination of real and synthetic data. Test data comprised real ultrasound data from Dataset A (paired t-test <0.05). \textbf{Bold text indicates the best results achieved}.}
    \centering
    \begin{tabular}{lcccccc}
        \toprule
        &Train Dataset &Accuracy &Sensitivity &Specificity &Precision &F1 \\
        \midrule
        \multirow{2}{*}{DenseNet121}
        &{Real-ultrasound only} &87.5&0.87&0.92	&0.88 &0.87 \\
        &{Real+50\%-High-contrast} &\textbf{90}	&\textbf{0.90}	&\textbf{0.94}&\textbf{0.89}&\textbf{0.89}	 \\
        \hline
        \multirow{2}{*}{ResNet34}
        &{Real-ultrasound only} &79.17	&0.77&0.87&0.81&0.79 \\
        &{Real+50\%-Stylized} &\textbf{87.5}	&\textbf{0.88}	&\textbf{0.93}	&\textbf{0.85}	&\textbf{0.86} \\
        \hline   
        \multirow{2}{*}{Squeezenet1.1}
        &{Real-ultrasound only} &72.5	&0.68	&0.86	&0.67	&0.67	\\
        &{Real+50\%-High-contrast} &\textbf{82.5}  &\textbf{0.80} &\textbf{0.89} &\textbf{0.82} &\textbf{0.81} 	\\
        \hline            
 \end{tabular}

\label{original_syn}
\end{table*}

\noindent As a subsequent procedure, we generate images in three varying sizes: 50\%, 100\%, and 200\% of the Dataset A test data size(545), accompanied by nine adjectives and a scenario devoid of any adjective. These images are subsequently integrated with the real image train data from Dataset A to establish a new training dataset, which is employed for training the classifiers. The validation and test datasets consistently comprise real ultrasound images from Dataset A, as elucidated in the 'Dataset' section. In the first row of Table \ref{table_adjectives}, the absence of an entry ("") indicates that no adjective was utilized during training. As demonstrated in Table\ref{table_adjectives}, where we display the average accuracy, the utilization of adjectives consistently outperforms scenarios without adjectives in the majority of cases. Table \ref{original_syn} presents quantitative results comparing the performance of classifiers when using real ultrasound data only versus the best synthetic datasets generated from the optimal performance of each classifier as shown in Table \ref{table_adjectives}. Our proposed pipeline demonstrates superior performance over the original datasets in terms of accuracy, sensitivity, specificity, precision, and F1 score (Table \ref{original_syn}) (paired t-test <0.05).

\noindent\textbf{Selection of adjectives.} In our pipeline, a common question pertains to the selection of adjectives. While there is inherent randomness in the generation process, we sought to elucidate the selection of adjectives through experiments.

\begin{figure}[h]
    \centering
    \begin{minipage}{0.5\textwidth}
        \centering
        \subfloat[]{\includegraphics[width=\textwidth]{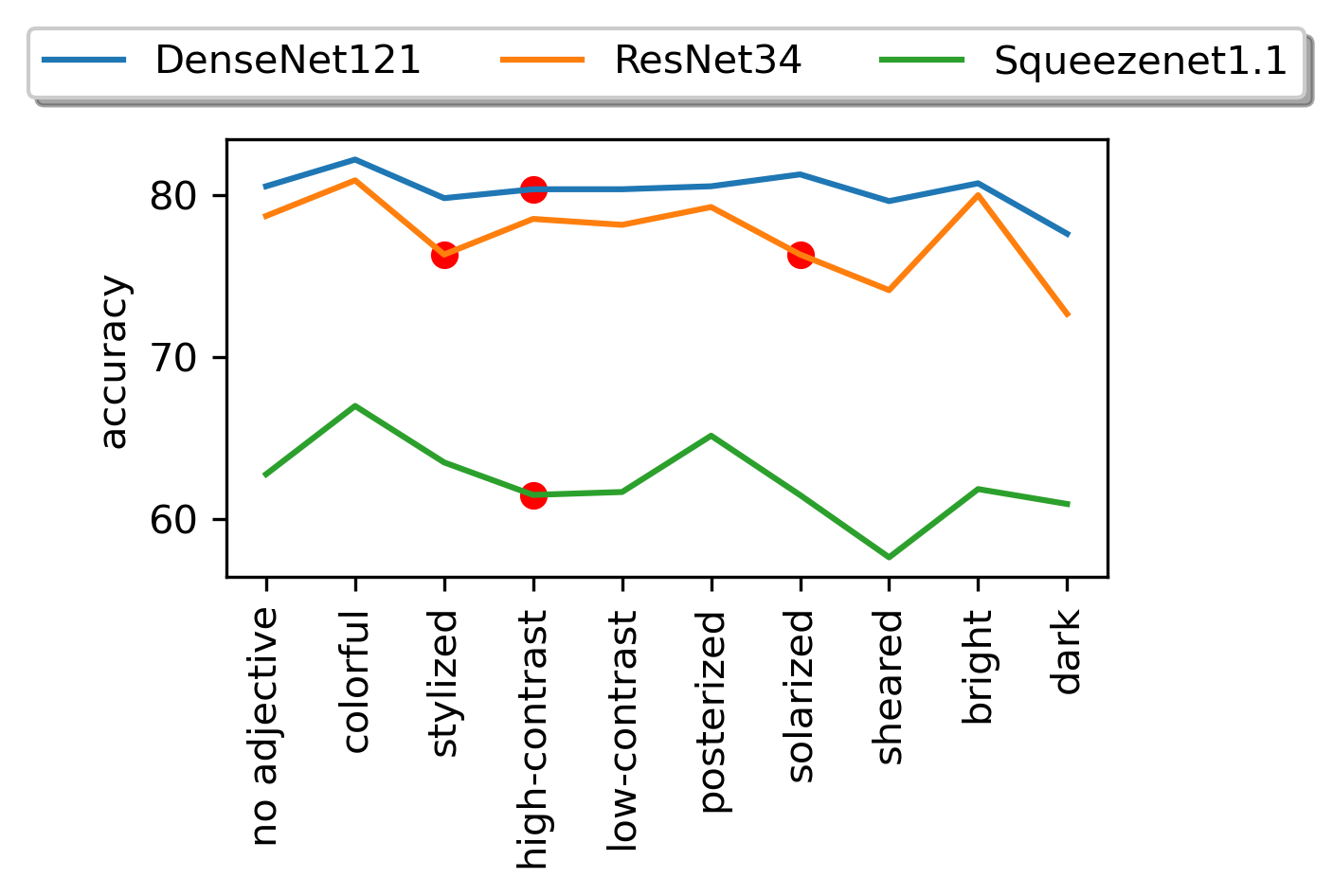}}
    \end{minipage}%
    \begin{minipage}{0.5\textwidth}
        \centering
        \subfloat[ ]{%
            \begin{tabular}{cc}
                \includegraphics[width=0.35\textwidth]{figures/benign_revised.png} &
                \includegraphics[width=0.35\textwidth]{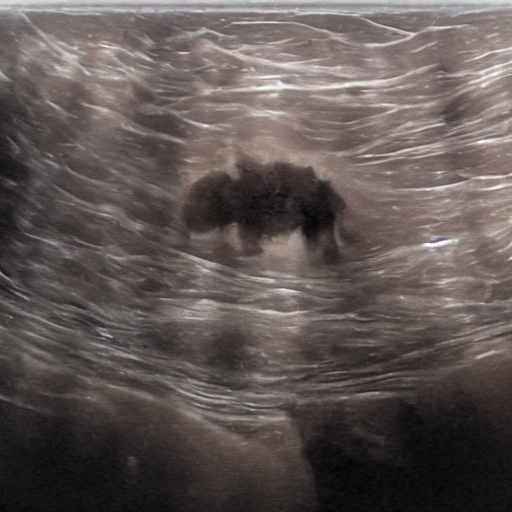} \\
                \includegraphics[width=0.35\textwidth]{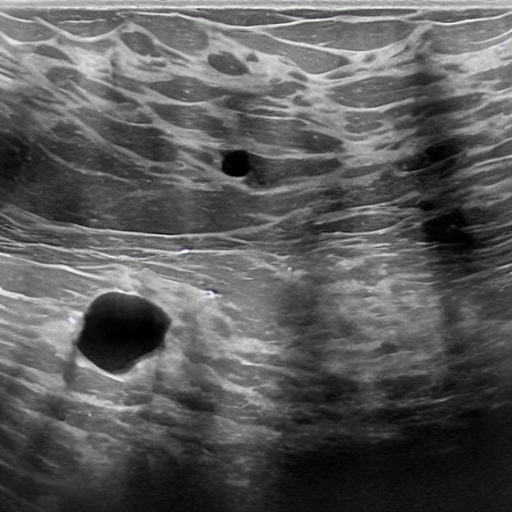} &
                \includegraphics[width=0.35\textwidth]{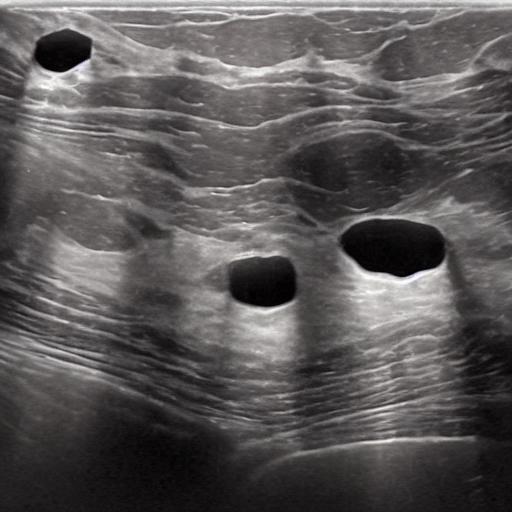}
            \end{tabular}
        }
    \end{minipage}
    \caption{(a) Accuracy results when networks are trained on real data only and tested on synthetic datasets only. (b) Some examples of synthesized images. Top-left:  Original B-mode ultrasound data, synthesized images using top-right: Colorful, bottom-left: Solarized, bottom-right: High-contrast adjectives.  }
    \label{adjectives_plot}
\end{figure}

\noindent By testing synthetic datasets with various adjectives on classifiers trained on the real ultrasound train set only, we aimed to shed light on this aspect. Figure \ref{adjectives_plot}-(a) illustrates the outcomes, revealing that the best adjectives (depicted by red circles) do not consistently yield the highest or lowest performance. Notably, achieving the best accuracy implies that the generated images closely resemble the real ultrasound data, potentially indicating that the SD model is lacking new information(diversity). Conversely, the worst performance manifests that the generated images are distant from the distribution of the real ultrasound data. In contrast, we aim to guide existing diffusion models towards a broader comprehension of classes by emphasizing the distinguishing characteristics necessary for discrimination, all while harnessing the full diversity of the underlying generative model. Figure \ref{adjectives_plot}-(b) shows qualitative results obtained using a sample of adjectives. 

\noindent\textbf{Effect of our proposed method in robustness.}
In this section, we assess the efficacy of our method in accurately classifying unseen datasets, specifically Dataset B \cite{Dataset_c}. Table \ref{robustness_table} illustrates that our top-performing adjectives, identified from Table \ref{adjectives_plot} for each classifier, consistently outperform the classifiers trained with real data only across all scenarios.

\begin{table*}[h]
\caption{Accuracy results evaluating the robustness of our pipeline on unseen test Dataset B (paired t-test <0.05). \textbf{Bold text indicates the best results achieved.}}
    \centering
    \begin{tabular}{lc|c|c}
        \toprule
        &Train Dataset &Adjective & Test Set ‌‌B Accuracy \\
        \midrule
        \multirow{2}{*}{DenseNet121}
        &{Real-ultrasound only}&-  &79.66\\
        &{Real+50\% Augmented} &{"High-Contrast"} &\textbf{83.22}\\
        \hline
        \multirow{2}{*}{ResNet34}
        &{Real-ultrasound only}&-  &77.44  \\
        &{Real+50\% Augmented} &{"Stylized"}  &\textbf{80.44}\\
        \hline   
        \multirow{2}{*}{Squeezenet1.1}
        &{Real-ultrasound only}&- & 64.44\\
        &{Real+50\% Augmented} & {"High-Contrast"}  & \textbf{72.89} \\
        \hline            
 \end{tabular}

\label{robustness_table}
\end{table*}
\begin{figure}[h]
    \centering
    {{\includegraphics[width=7cm]{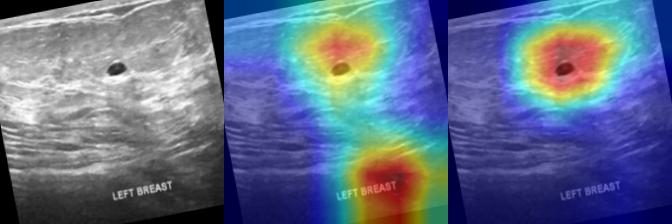} }}

    {{\includegraphics[width=7cm]{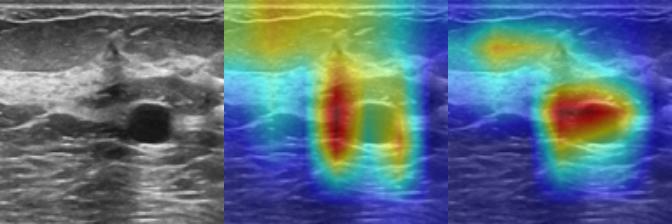} }}
    \caption{Heat map results depict classification network attention. Columns from right to left: Real ultrasound data, heat map output when classification network (Resnet34 top, DenseNet121 bottom) was trained solely on real ultrasound dataset, and heat map output when classification network (Resnet34 top, DenseNet121 bottom) was trained on real ultrasound dataset augmented with 50\% "Stylized" adjective (top) and 50\% "High-Contrast" adjective (bottom).}%
    \label{robustness_image}%
\end{figure}

\noindent Analyzing the heat-map images depicted in Figure \ref{robustness_image}, we note that models trained on original datasets primarily focus on textual elements (part a) and the ultrasound image's shadow (part b) rather than tumors. The classifier's focus on the text exemplifies a typographic attack \cite{goh2021multimodal}, where the classifier prioritizes textual elements over the depiction of tumors in the ultrasound image. In contrast, we observe that integrating synthetic datasets with real data enhances the reliability of the decision-making process.

\section{Conclusion}
\justify This study proposes a novel pipeline for expanding small-scale ultrasound datasets by incorporating informative images rather than simply increasing dataset size. This approach not only enhances classification accuracy but also bolsters the network's robustness and reliability by diversifying the dataset. Our proposed method addresses the challenges associated with training diffusion models, which typically demand substantial memory resources, thereby facilitating broader access for fine-tuning this extensive model to address various problems.

\bibliographystyle{splncs04}
\bibliography{ref}

\begin{thebibliography}{10}
\providecommand{\url}[1]{\texttt{#1}}
\providecommand{\urlprefix}{URL }
\providecommand{\doi}[1]{https://doi.org/#1}

\bibitem{intrinsic}
Aghajanyan, A., Zettlemoyer, L., Gupta, S.: Intrinsic dimensionality explains the effectiveness of language model fine-tuning. arXiv:2012.13255 [cs]  (2020)

\bibitem{Dataset}
Al-Dhabyani, W., Gomaa, M., Khaled, H., Fahmy, A.: Dataset of breast ultrasound images. Data in Brief  (2020)

\bibitem{ali2022spot}
Ali, H., Murad, S., Shah, Z.: Spot the fake lungs: Generating synthetic medical images using neural diffusion models. In: Irish Conference on Artificial Intelligence and Cognitive Science. pp. 32--39. Springer (2022)

\bibitem{alsinan2020gan}
Alsinan, A.Z., Rule, C., Vives, M., Patel, V.M., Hacihaliloglu, I.: Gan-based realistic bone ultrasound image and label synthesis for improved segmentation. In: Medical Image Computing and Computer Assisted Intervention--MICCAI 2020: 23rd International Conference, Lima, Peru, October 4--8, 2020, Proceedings, Part VI 23. pp. 795--804. Springer (2020)

\bibitem{compvis2023stable}
{CompVis}: {Stable Diffusion V1.4 - Original} (2023), \url{https://huggingface.co/CompVis/stable-diffusion-v-1-4-original}, accessed: [Insert Date of Access]

\bibitem{Dataset_c}
Ultrasound breast images for breast cancer, \url{https://www.kaggle.com/datasets/vuppalaadithyasairam/ultrasound-breast-images-for-breast-cancer/data}

\bibitem{dhariwal2021diffusion}
Dhariwal, P., Nichol, A.: Diffusion models beat gans on image synthesis. Advances in neural information processing systems  \textbf{34},  8780--8794 (2021)

\bibitem{fujioka2020utility}
Fujioka, T., Mori, M., Kubota, K., Oyama, J., Yamaga, E., Yashima, Y., Katsuta, L., Nomura, K., Nara, M., Oda, G., et~al.: The utility of deep learning in breast ultrasonic imaging: a review. Diagnostics  \textbf{10}(12), ~1055 (2020)

\bibitem{goh2021multimodal}
Goh, G., Cammarata, N., Voss, C., Carter, S., Petrov, M., Schubert, L., Radford, A., Olah, C.: Multimodal neurons in artificial neural networks. Distill  \textbf{6}(3), ~e30 (2021)

\bibitem{resnet}
He, K., Zhang, X., Ren, S., Sun, J.: Deep residual learning for image recognition. In: Proceedings of the IEEE conference on computer vision and pattern recognition. pp. 770--778 (2016)

\bibitem{ho2020denoising}
Ho, J., Jain, A., Abbeel, P.: Denoising diffusion probabilistic models. Advances in neural information processing systems  \textbf{33},  6840--6851 (2020)

\bibitem{LORA}
Hu, E.J., Shen, Y., Wallis, P., Allen-Zhu, Z., Li, Y., Wang, S., Wang, L., Chen, W.: Lora: Low-rank adaptation of large language models. arXiv:2106.09685v2 [cs.CL]  (2021)

\bibitem{huang2023style}
Huang, B., Xu, Z., Chan, S.C., Liu, Z., Wen, H., Hou, C., Huang, Q., Jiang, M., Dong, C., Zeng, J., et~al.: A style transfer-based augmentation framework for improving segmentation and classification performance across different sources in ultrasound images. In: International Conference on Medical Image Computing and Computer-Assisted Intervention. pp. 44--53. Springer (2023)

\bibitem{densenet}
Huang, G., Liu, Z., van~der Maaten, L., Weinberger, K.Q.: Densely connected convolutional networks. In: Proceedings of the IEEE Conference on Computer Vision and Pattern Recognition (2017)

\bibitem{SqueezeNet}
Iandola, F.N., Han, S., Moskewicz, M.W., Ashraf, K., Dally, W.J., Keutzer, K.: Squeezenet: Alexnet-level accuracy with 50x fewer parameters and $<$0.5mb model size. arXiv:1602.07360  (2016)

\bibitem{kaggle}
Kaggle, \url{https://www.kaggle.com/docs/datasets}

\bibitem{katakis2023generation}
Katakis, S., Barotsis, N., Kakotaritis, A., Tsiganos, P., Economou, G., Panagiotopoulos, E., Panayiotakis, G.: Generation of musculoskeletal ultrasound images with diffusion models. BioMedInformatics  \textbf{3}(2),  405--421 (2023)

\bibitem{kazerouni2022diffusion}
Kazerouni, A., Aghdam, E.K., Heidari, M., Azad, R., Fayyaz, M., Hacihaliloglu, I., Merhof, D.: Diffusion models for medical image analysis: A comprehensive survey. arXiv preprint arXiv:2211.07804  (2022)

\bibitem{khader2022medical}
Khader, F., Mueller-Franzes, G., Arasteh, S.T., Han, T., Haarburger, C., Schulze-Hagen, M., Schad, P., Engelhardt, S., Baessler, B., Foersch, S., et~al.: Medical diffusion--denoising diffusion probabilistic models for 3d medical image generation. arXiv preprint arXiv:2211.03364  (2022)

\bibitem{lee2016digital}
Lee, C.I., Lehman, C.D.: Digital breast tomosynthesis and the challenges of implementing an emerging breast cancer screening technology into clinical practice. Journal of the American College of Radiology  \textbf{13}(11),  R61--R66 (2016)

\bibitem{liang2022sketch}
Liang, J., Yang, X., Huang, Y., Li, H., He, S., Hu, X., Chen, Z., Xue, W., Cheng, J., Ni, D.: Sketch guided and progressive growing gan for realistic and editable ultrasound image synthesis. Medical Image Analysis  \textbf{79},  102461 (2022)

\bibitem{litjens2017survey}
Litjens, G., Kooi, T., Bejnordi, B.E., Setio, A.A.A., Ciompi, F., Ghafoorian, M., Van Der~Laak, J.A., Van~Ginneken, B., S{\'a}nchez, C.I.: A survey on deep learning in medical image analysis. Medical image analysis  \textbf{42},  60--88 (2017)

\bibitem{rombach2021highresolution}
Rombach, R., Blattmann, A., Lorenz, D., Esser, P., Ommer, B.: High-resolution image synthesis with latent diffusion models (2021)

\bibitem{FID}
Seitzer, M.: {pytorch-fid: FID Score for PyTorch}. \url{https://github.com/mseitzer/pytorch-fid} (August 2020), version 0.3.0

\bibitem{stevens2024dehazing}
Stevens, T.S., Meral, F.C., Yu, J., Apostolakis, I.Z., Robert, J.L., Van~Sloun, R.J.: Dehazing ultrasound using diffusion models. IEEE Transactions on Medical Imaging  (2024)

\bibitem{stojanovski2023echo}
Stojanovski, D., Hermida, U., Lamata, P., Beqiri, A., Gomez, A.: Echo from noise: synthetic ultrasound image generation using diffusion models for real image segmentation. arXiv preprint arXiv:2305.05424  (2023)

\bibitem{svensson2010advanced}
Svensson, W.E., Stewart, V.R.: Advanced applications of breast ultrasound. Breast Cancer pp. 70--98 (2010)

\bibitem{yap2017automated}
Yap, M.H., Pons, G., Marti, J., Ganau, S., Sentis, M., Zwiggelaar, R., Davison, A.K., Marti, R.: Automated breast ultrasound lesions detection using convolutional neural networks. IEEE journal of biomedical and health informatics  \textbf{22}(4),  1218--1226 (2017)

\bibitem{zhao2022ood}
Zhao, B., Yu, S., Ma, W., Yu, M., Mei, S., Wang, A., He, J., Yuille, A., Kortylewski, A.: Ood-cv: a benchmark for robustness to out-of-distribution shifts of individual nuisances in natural images. In: European Conference on Computer Vision. pp. 163--180. Springer (2022)

\end{thebibliography}

\end{document}